\documentstyle[12pt, aaspp4]{article}

\begin{document}

\title{Decomposition of the Superwind in M82\footnote{Based on data
collected at Subaru Telescope, which is operated by the National
Astronomical Observatory of Japan.}}

\author{Youichi Ohyama$^2$, Yoshiaki Taniguchi$^3$, Masanori Iye$^{4,5}$,
Michitoshi Yoshida$^6$, Kazuhiro Sekiguchi$^2$, Tadafumi Takata$^2$,
Yoshihiko Saito$^4$, Koji S. Kawabata$^4$, Nobunari Kashikawa$^4$,
Kentaro Aoki$^7$, Toshiyuki Sasaki$^2$, George Kosugi$^2$, Kiichi
Okita$^2$, Yasuhiro Shimizu$^6$, Motoko Inata$^4$, Noboru Ebizuka$^8$,
Tomohiko Ozawa$^9$, Yasushi Yadomaru$^9$, Hiroko Taguchi$^{10}$, and
Ryo Asai$^{11}$}

\affil{$^2$Subaru Telescope Office, National Astronomical Observatory
               of Japan, 650 N. A`ohoku Place, University Park, Hilo,
               HI 96720, USA \\
           $^3$Astronomical Institute, Graduate School of Science,
               Tohoku University, Aramaki, Aoba, Sendai 980-8578, Japan\\
           $^4$National Astronomical Observatory of Japan, 2-21-1 Osawa,
               Mitaka, Tokyo 181-8588, Japan \\
           $^5$Department of Astronomy, Graduate University for Advanced
                Studies, 2-21-1, Osawa, Mitaka, Tokyo 181-8588, Japan \\
           $^6$Okayama Astrophysical Observatory, National Astronomical
               Observatory of Japan, Kamogata-cho, Asakuchi-gun, Okayama
               719-0232, Japan \\
           $^7$Institute of Astronomy, Graduate School of Science,
               University of Tokyo, Mitaka 181-1015, Japan \\
           $^8$Cosmic Radiation Laboratory, Institute of Physical and
               Chemical Research (RIKEN), Wako, Saitama, 351-0198, Japan \\
           $^9$Misato Observatory, 180 Matsugamine, Misato-cho,
               Amakusa-gun, Wakayama, 640-1366, Japan\\
           $^{10}$Department of Astronomy and Earth Sciences, Tokyo Gakugei
               University, 4-1-1 Nukui-Kitamachi, Koganei, Tokyo 184-8501,
	       Japan \\
	   $^{11}$Systems Engineering Consultants Co., Ltd., Shibuya-ku,
		Tokyo 150-0031, Japan}

\begin{abstract}

We present new optical images ($B$, $V$, and H$\alpha$) of the
archetypical starburst/superwind galaxy M82 obtained with the 8.2 m
Subaru Telescope to reveal new detailed structures of the
superwind-driven nebula and the high-latitude dark lanes.
The emission-line nebula is decomposed into (1) a ridge-dominated
component comprising numerous filament/loop sub-structures whose
overall morphology appears as a pair of narrow cylinders, and
(2) a diffuse component extended over much wider opening angle from
the nucleus.
We suggest that these two components have different origins.
The ridge-dominated component appears as a pair of cylinders rather
than a pair of cones. Since this morphological property is similar
to that of hot plasma probed by soft X-ray, this component seems to
surround the hot plasma.
On the other hand, the diffuse component may arise from dust grains
which scatter stellar light from the galaxy.
Since inner region of this component is seen over the prominent
^^ ^^ X"-shaped dark lanes streaming out from the nuclear region
and they can be reproduced as a conical distribution of dust grains,
there seems to be a dusty cold outflow as well as the hot one probed by
soft X-ray and shock-excited optical emission lines.
If this is the case, the presence of such high-latitude
dust grains implies that neutral gaseous matter is also
blown out during the course of the superwind activity.
\end{abstract}


\section{INTRODUCTION}

Bursts of massive star formation (starbursts) are considered to play
important roles in the evolution of galaxies from high redshift to the
present day.
Approximately 10\% of nearby galaxies show evidence for such starburst
activity (Weedman et al. 1981; Balzano 1983; Ho et al. 1997).
In particular, some of them also show evidence for galactic-scale
bipolar outflows (superwinds) driven by collective effects of a large
number of supernova explosions occurred in the central region of
galaxies (e.g., Chevalier \& Clegg 1985; Tomisaka \& Ikeuchi 1988;
Heckman et al. 1990; Suchkov et al. 1994; Tenorio-Tagle \& Munoz-Tunon
1998; Strickland \& Stevens 2000).
The superwind phenomenon is also very important as well as the starburst
activity itself because it provides metal-enriched gas into
the intergalactic medium.

Since the identification of optical emission-line filaments in M82
(Lynds \& Sandage 1963; Burbidge et al. 1964), much attention has been
paid to this galaxy and it is now widely accepted that M82 is the
archetypal starburst galaxy with evidence for the bipolar superwinds
emanating from the central region to the outer halo area (e.g., Heckman
et al. 1987; Nakai et al. 1987; Bland \& Tully 1988; Shopbell \&
Bland-Hawthorn 1998; Lehnert et al. 1999).
Because of its proximity (a distance to M82, $D=$ 3.63 Mpc;
Freedman et al. 1994), M82 allows us to investigate detailed
characteristics of the superwind activity.
Indeed, previous studies of the M82 superwind have shown that the gas
in this superwind takes various phases:
I) hot ($T \sim 10^7$ K) soft X-ray emitting gas (e.g., Bregman,
Schulman, \& Tomisaka 1995; Strickland, Ponman, \& Stevens 1997;
Lehnert et al. 1999), II) warm ($T \sim 10^4$ K) ionized gas probed by
optical emission lines (e.g., Heckman et al. 1987; Bland \&Tully 1988;
McKeith et al. 1995; Shopbell \& Bland-Hawthorn 1998), III) cool ($T
\sim 100$ K) neutral gas probed by dust grains, including
i) dusty molecular-gas halo (e.g., Scarrott, Eaton, \& Axon 1991;
Sofue et al. 1992) and ii) complex dark lanes seen in the upper disk
regions (Ichikawa et al. 1994; 1995), and IV) cold ($T \sim 10 - 100$ K)
molecular and atomic gas (e.g., Nakai et al. 1987; Yun, Ho, \& Lo 1993).
However, it is not yet clear how such various gaseous components are
related to each other although the observational properties of
each component have been discussed extensively.

In order to improve our understanding of this superwind activity, we
present new sensitive high-resolution images of M82 taken with the 8.2
m Subaru telescope (Kaifu et al. 1998).
These images are analyzed in various ways to reveal new detailed
structures of the galactic disk (stars), dark lanes (dust grains),
ridge-dominated and diffuse components (shock-heated warm gas and dust
grains scattering the nuclear light, respectively) of the nebula.
Combining with the previous information of X-ray image for hot ionized
gas, we investigate the multi-structured superwind phenomenon in M82.

\section{OBSERVATIONS AND DATA ANALYSIS}

The optical imaging observations of M82 were made with the FOCAS
(Faint Object Camera and Spectrograph: Kashikawa et al. 2000, Yoshida et al.
2000) on the
Subaru Telescope during a commissioning run on 2000 February 2 (UT).
The seeing condition was 0.7\arcsec ~ - 0.8\arcsec~ for all images.
In order to investigate the emission-line nebula associated with the
superwind, we obtained direct images with two narrowband filters, N658
and N642.
The N658 filter centered at 6588\AA~ with a band width of 73\AA~ was
used to probe both H$\alpha$ $\lambda$6563 and [N {\sc ii}]
$\lambda\lambda$6548, 6583 emission lines (hereafter, H$\alpha$ + [N
{\sc ii}]), while the N642 filter centered at 6428\AA~ with a band
width of 127\AA~ was used to measure the adjacent continuum emission.
We obtained five N658 images and three N642 ones.
Each exposure time was set to be 120 seconds and thus the total
integration was 600 seconds and 360 seconds for the N658 and N642
images, respectively.
We also obtained optical $B$- and $V$-band images.
The total integration time was 120 seconds (4 $\times$ 30 seconds) and
75 seconds (3 $\times$ 25 seconds) for $B$ and $V$, respectively.
No photometric standard star was observed and thus we do not calibrate
the flux scale of any images.
All the data reduction was made using the dedicated software for FOCAS
implemented with IDL (Yoshida et al. 2000) and IRAF\footnote
{IRAF is distributed by the
National Optical Astronomy Observatories, which are operated by the
Association of Universities for Research in Astronomy, Inc., under
cooperative agreement with the National Science Foundation.}.

These images were further processed in the following ways.
A pure H$\alpha$ + [N {\sc ii}] emission-line image (hereafter
H$\alpha$ + [N {\sc ii}] image) was made by subtracting the scaled
N642 image from the N658 one
\footnote
{
The scaling of the N642 image was done by comparing fluxes of nearby stars around M82.
Although we did not take account of the possible underestimate of the scaled N642 flux due to H$\alpha$ absorption in N658 image, this effect is estimated to be negligibly small since the disk stellar emission could be subtracted clearly over the entire disk.
}
.
A pure continuum image around 6500\AA~ (hereafter 6500\AA~ continuum
image) was made by summing N642 and N658 images and subtracting the
H$\alpha$ + [N {\sc ii}] image from the summed one.
The H$\alpha$ + [N {\sc ii}] image is shown superimposed on the
true-color images of the M82 galaxy disk (made from $B$, $V$, and
6500\AA~ continuum images) in Figure 1.
The emission-line nebula appears to consist of the following two
major components; (1) the ridge component which corresponds to
the bipolar filament-dominated component,
and (2) the diffuse component which is more widely
extended around the ridge component.

In the ridge component (Figure 2), many shell- and loop-like filaments are
recognized.
Their typical sizes are 0.8\arcsec~ (unresolved) 
- 2\arcsec~ wide and 5\arcsec~ - 20\arcsec~ long.
In order to show such filaments more clearly, we developed a ridge
detection program which searches for any regions with large negative
value of the second derivative of the image (i.e., tracing local peaks
in the emission-line image).
The sky area without filaments was extracted from the H$\alpha$ + [N {\sc
ii}] image by masking out the ridges detected, and the image of the
diffuse component was made by filling-in the masked regions by
interpolating the surrounding pixel values.
Subtracting the diffuse component image from the H$\alpha$ + [N {\sc
ii}] image, we obtain an image of the ridge component.
Figure 3 shows images of H$\alpha$ + [N {\sc ii}] nebula shown
separately in ridge and diffuse components.
Note that since our ridge-detection program can only detect the extreme limbs
or tangent points of the structures, structures resolved slightly around
the peak of the ridges are included in the diffuse component, and are notable
especially near the root of the nebula.
Therefore the amount of the diffuse component is overestimated.
The broadband color image was made by dividing the $B$-band image with
the 6500\AA~ continuum image, and is shown in Figure 4 along with
the ridge and diffuse components of the nebula.

\section{RESULTS and DISCUSSION}

\subsection{The Ridge Component}

The ridge component comprises many filaments along the superwind.
Each filament shows either shell- or loop-like structures directing
radially from the nuclear region.
Similar filament-dominated emission-line nebulae are also found in some
well-known Galactic objects with an outflow [such as M1-67: a
Wolf-Rayet star with stellar wind (Grosdidier et al. 2001) and M57: a
planetary nebula with a ring (Komiyama et al. 2000)].
It is widely accepted that such a filament can
be a local shock front between the inner expanding matter and the
outer interstellar medium. The observed filaments in M82 are
also considered to be driven by shocks. This interpretation is supported
by the optical emission-line diagnostics (Heckman et al. 1987)
although their data probe both ridge and diffuse components.

As mentioned before,
the overall structure of the ridge component looks like a pair of
elongated cylinders rather than a pair of cones, and hence hereafter
we simply denote it as the cylinder.
The diameter of the cylinder at the galaxy disk plane (35\arcsec~
- 40\arcsec)
is similar to, or slightly larger than, that of the nuclear
starburst region traced both by near-infrared emission (30\arcsec~ 
- 35\arcsec;
e.g., Lester et al. 1990) and by an ensemble of non-thermal point-like
sources (e.g., Kronberg, Biermann, \& Schwab 1985).
It seems worthwhile noting that the cylinder shows the sharp boundary,
in particular at NE, SE, and SSE of the nebula, causing the limb-enhanced
morphology.
All these features can be well reproduced by numerical
simulations of the superwind in which radiative shock occurs around the
expanding hot gas (e.g., Suchkov et al 1994; Strickland \& Stevens 2000)
(see section 3.3).

The hot gaseous component probed by soft X-ray emission is
known to be collimated rather tightly along the minor axis of the
galaxy than the optical emission-line nebula (Bregman et al. 1995;
Strickland et al. 1997; Lehnert et al. 1996).
We found that the ridge component surrounds the hot gaseous nebula more
closely than the diffuse component. However, the tangential profiles
along the galaxy disk at high latitude look different between the
ridge component (showing limb-enhanced profile) and the soft X-ray nebula
(showing the ridge along
the centerline).
Note, however, that such difference can be explained if we assume that
the optical filaments are distributed on the thin surface around the
inner hot gas component.
All these lines of evidence suggest that the radiative shock at the
thin interface between the expanding hot X-ray emitting gas and the
ambient cold gaseous matter would be responsible for the ridge component.

\subsection{The Diffuse Component}

The diffuse component was first found by Bland \& Tully (1988) and was
confirmed by Shopbell \& Bland-Hawthorn (1998).
It shows rather oval shape elongated along the minor axis of the
galaxy as a whole, but is divided into NW and SE parts around the disk plane.
It is extended more widely in the tangential direction (parallel to
the galaxy disk) and does not show a limb-enhanced morphology, being
much different from the ridge component
\footnote{
As noted in the earlier section, separating the ridge and diffuse components by the ridge-detection program could not be perfectly done because some part of the flux separated as ``diffuse'' may actually be structures which are resolved slightly around the peak of the ridges.
However, since the ridge component sppares to be a collimated cylindrical structure outer parts of the nebula around the ridge component are likely to be truly ``diffuse''.
Therefore the comparison of the overall shapes between the ridge and diffuse components can be made with less ambiguity.
}
.
These observational properties suggest that this component does not
have a form of thin surface but occupies a large part of the volume
probed by optical emission lines.

Here a question arises as what the origin of the diffuse component is.
Previous imaging polarimetries of the M82 nebula have revealed
strongly-polarized emission whose $E$ vectors show circular symmetric pattern
around the nucleus. Since this property is a typical signature of the
reflection nebulae, the dust scattering of the galaxy light would contribute
to the nebula (e.g., Bingham et al. 1976; Schmidt, Angel, \& Cromwell 1976).
It is thus suggested that the diffuse component arises from
the scattering of the galaxy light at dusty halo.

In addition to the dusty halo, M82 is well known to have rich dark lanes
across the galaxy disk.
They are peculiar and complex in shape, and are remarkably
different from those found in normal spiral galaxies.
The most prominent and peculiar dark lane is seen on SE side of the disk
(the front side of the disk: e.g., Shopbell \& Bland-Hawthron 1998; McKeith
et al. 1995).
It runs through the nuclear region out to 2.5 arcmin (2.6
kpc) from the nucleus (Figure 5).
Interestingly it curves toward both SE and SW directions at both ends of
the dark lane, and can be traced up to $\sim 1$\arcmin~ ($\sim 1$ kpc
in projection) above/below the galaxy disk at their tips.
Similar but less-prominent dark lanes are also found at another side of the
disk emanating to NE and NW from the nuclear region.
These two sets of the dark lanes make an ``X''-shaped morphology around the
nucleus. They are considered to be distributed not within the galaxy
disk but above/below the disk (e.g., Ichikawa et al. 1994; 1995).
We find that the diffuse component can be seen even on the dark lanes,
suggesting that dust grains responsible for the dark lanes are also
responsible for the scattering of the diffuse component.

The next question is addressed to the origin of the dark lanes.
The dark lanes have the following characteristics.
First, the X-shaped dark lanes seem to be streaming out from the nucleus.
Second, the NW boundary (and also probably the NE boundary) of the SE
dark lane is smoothly curved comparing with other complicated dark lanes.
Finally one-dimensional profile of the SW dark lane, cut perpendicular
to the disk across the dark lane (Figure 5), shows highly asymmetric
structure of the extinction, i.e., the dust extinction increases
monotonically toward the NW side, and then decreases suddenly
at the NW edge the SE dark lane.
Such characteristics can be understood if we assume that the dust is
distributed in a wide cone-like geometry and we see a higher
extinction near the cone surface due to projection effect.
This supports a scenario of the dusty superwind in which dust grains are
provided from the nuclear region with the outflow associated with the
superwind probed in optical emission-lines and soft X-ray emission.
If this is the case, high-latitude dusty halo may be created in a similar way
as for the dark lanes (i.e., the dusty superwind) since there are no
distinct boundaries in the diffuse component dividing the dust scattering
on the dark lane near galaxy plane and high-latitude halo area up to
$\gtrsim 2$ kpc from the disk.
Yun et al. (1993) suggested that a part of the high-latitude HI gas of M82
may be affected dynamically by the superwind, supporting for the presence of a
high-latitude cold gas associated with the superwind outflow.
Although we do not rule out an alternative
possibility that the past tidal interaction with the neighbor galaxy
M81 has brought some dusty matter from the galaxy disk up to the halo
(Yun, Ho, \& Lo 1993; 1994), the observational properties discussed
in this paper suggest that the dusty superwind scenario is more reasonable.

Such a dusty cold-gas outflow can be made as a result of the pushing
out and/or dragging out process of the dusty gas around the nucleus by the
expanding hot high-pressure gas within the superwind (e.g., Suchkov et
al. 1994; Strickland \& Stevens 2000).
Indeed dusty outflows associated with superwinds have been suggested by
optical observations (Nakai 1989; Heckman et al. 2000; Veilleux \& Rupke
2002; Rupke, Veilleux, \& Sanders 2002)
and submillimeter observations (Alton, Davies, \& Bianchi 1999).

Finally we give a comment on the previous claim that even the inner
filamentary nebula shows strong polarization whose direction
is nearly aligned to that of
the outer diffuse nebula (Scarrott et al. 1991), which could be inconsistent
with our conclusion since no polarized flux is expected from the
shock-excited emission.
We suspect that the imaging polarimetry with insufficient spatial resolution
could be the cause of this apparent contradiction. Separate polarimetric
measurement of the filaments in
  the ridge and diffuse components are required to
disentangle the polarimetric properties of the
narrow filamentary structures (typically $\lesssim 2$\arcsec~ in
width) and the overlapping diffuse component.

\subsection{Implications for Superwind Model}

Important characteristics of the M82 superwind nebula are that
the shock-driven ridge component contains numerous shells and filaments
and its overall shape is rather collimated out
to outer region of the nebula.
In recent hydrodynamical simulations of the superwind conducted by Strickland \& Stevens (2000), two types of the interstellar medium (ISM) density distribution, thin or thick disk, are adopted to see its effect on the evolution of the wind.
It turns out that the model with the thick ISM can reproduce the characteristics of M82 nebula:
The thick and dense wall of ISM created around the starbursting region works to collimate the wind, and Rayleigh-Taylor instability at the wall creates small-scale structures which are eventually dragged out with the wind to form shells/filaments (Strickland \& Stevens 2000).
Although these hydrodynamical simulations with thick ISM look successful, we point out the following points which require some cautions.
As Strickland \& Stevens (2000) noted, the thick ISM cannot reproduce the observed rotation curve of M82.
Therefore one may need more realistic ISM distribution models or improved treatment of gas dynamics at the circumnuclear region to reproduce both a collimated wind and the rotation curve consistently.
It seems also noteworthy that the morphology of the small-scale structures of the nebula depend rather strongly on the numerical resolution in the simulation, and higher resolution tends to result in finer shell/filament structures (Strickland \& Stevens 2000).
Therefore one may need even finer-scale numerical simulation to enable us to compare sizes and spatial distributions of shells/filaments between observations and the models.
Note that, although observed small-scale structures with a typical size of $90 - 350$ pc (5\arcsec~ - 20\arcsec) are already resolved by their simulations with a cell size of $4.9 - 14.6$ pc, the simulation resolution would not be fine enough to resolve the circumnuclear region where smaller-scale instability begins to work.
Finally, we show a cartoon in which all the superwind components discussed here are summarized (Figure 6).

\acknowledgments
We would like to thank all staffs of the Subaru
telescope.

\newpage

\newpage

\figcaption{True color image of M82 made from $B$, $V$, and 6500\AA~
continuum images.
The total H$\alpha$ + [N {\sc ii}] image (ridge plus diffuse
components) is shown superimposed.
Some horizontal lines and circles are the artifacts in the image processing.
See a main text for the detail of the image processing.
}

\figcaption{
Some interesting small-scale structures within southeastern superwind nebula.
Central panel shows H$\alpha$ + [N {\sc ii}] image of the ridge component.
The ridge component image is rotated 65 degree clockwise so that southeastern nebula is seen in horizontal direction and the stellar disk is seen at vertical direction at the right edge of the panel.
Selected regions are marked with blue boxes, and each of them is labeled from A to F.
Six sub-images around the central one (labeled as A - F) show enlarged images of each corresponding sub-region marked in the central panel.
Total (ridge plus diffuse) component is shown in these panels to present full detail of the structures.
All images are shown in a same scale, and their box size is $30 \times 30$ arcsec$^2$ for all images except for the panel F whose size is $30 \times 60$ arcsec$^2$.
All images (central one and six sub-images) are shown in square-root intensity scale (as shown in lower right color bar), and maximum and minimum intensity cut levels are adjusted to present the fine structures clearly depending on the positions and the intensity of each structure.
}

\figcaption{H$\alpha$ + [N {\sc ii}] images decomposed into ridge and
diffuse components.
Upper three panels show entire nebula ($4.0\times 5.0$ arcmin$^2$) of
ridge component (left), diffuse component (middle), and total (ridge
plus diffuse) component (right).
These three images in the upper row are shown in logarithmic scale.
Lower three panels show inner $1.5\times 2.0$ arcmin$^2$ region of the
upper image.
Ridge, diffuse, and total components are shown separately as in the
same way as for the upper panels.
These three images in the lower panels are shown in square-root scale.
Tick marks are shown every 10\arcsec~ for all images.
}

\figcaption{Comparison of various components of at central $3.5\times
3.5$ arcmin$^2$ regions of M82.
(upper left) the 6500\AA~ continuum image representing the stellar
distribution along the galaxy disk shown in square root scale.
(upper right) the $B$-band image divided by the 6500\AA~ continuum
image representing the distribution of dark lanes shown in linear
scale.
(lower left) the ridge component of the emission-line nebula
representing the shock-driven nebula shown in logarithmic scale.
(lower right) the diffuse component of the emission-line nebula
representing the dust scattering component of the nebula shown in
logarithmic scale.
An outline of the X-shaped dark lanes (upper right) is superimposed
on the diffuse component.
Tick marks are shown every 10\arcsec~ for these four images.
}

\figcaption{One dimensional profile of the dark lane.
The cut region runs through 1\arcmin SWW of the nucleus along the disk,
and extendes over 1.5\arcmin~ perpendicular to the disk.
The profile is averaged over 10\arcsec~ along the disk.
The x-axis is shown in a relative distance from the mid-plane of the disk
in arcseconds, and y-axis is shown in a $B$-(red continuum) color in
magnitude with a tick corresponding to 0.1 mag.
See text for the red continuum.
}

\figcaption{Schematic view of M82 seen from the side.}


\begin{references}
\reference{1}{Alton, P. B., Davies, J. I., \&
               Bianchi, S. 1999, \aap, 343, 51}
\reference{1}{Balzano, V. A. 1983, ApJ, 268, 602}
\reference{1}{Bingham, R. G., McMullan, D., Pallister,
	W. S., White, C., Axon, D. J., \& Scarrott, S. M. 1976, \nat,
	259, 463}
\reference{1}{Bland, J., \& Tully, R. B. 1988, Nature, 334, 43}
\reference{1}{Bregman, J. N., Schulman, E., \& Tomisaka, K. 1995,
\apj, 439, 155}
\reference{1}{Burbidge, E. M., Burbidge, G. R., \& Rubin, V. C.
                  1964, ApJ, 140, 942}
\reference{1}{Chevalier, R. A., \& Clegg A. W. 1985, Nature, 317, 44}
\reference{1}{Freedman, W. L., et al. 1994, ApJ, 427, 628}
\reference{1}{Grosdidier, Y., Moffat, A. F. J., Blais-Ouellette,
               S. Joncas, G., \& Acler A. 2001, \apj, 562, 753}
\reference{1}{Heckman, T. M., Armus, L., \& Miley, G. K. 1987, \aj,
               93, 276}
\reference{1}{Heckman, T. M., Armus, L., \& Miley, G. K. 1990, \apjs,
               74, 833}
\reference{1}{Heckman, T. M., Lehnert, M. D., Strickland, D. K., \&
               Armus, L. 2000, ApJS, 129, 493}
\reference{1}{Ho, L. C., Filippenko, A. V., \& Sargent, W. L. W.
               1997, ApJ, 487, 579}
\reference{1}{Ichikawa, T., van Driel, W., Aoki, T., Soyano, T.,
         Tarusawa, K., \& Yoshida, S. 1994, \apj, 433, 645}
\reference{1}{Ichikawa, T., Yanagisawa, K., Itoh, N.,
         Tarusawa, K., van Driel, W., \& Ueno, M. 1995, \aj, 109, 2038}
\reference{1}{Kaifu, N. 1998, Proc. SPIE, 3352, 14}
\reference{1}{Kashikawa, N., et al. 2000, Proc. SPIE, 4008, 11}
\reference{1}{Komiyama, Y., Yagi, M., Miyazaki,
         S., et al. 2000, \pasj, 52, 93}
\reference{1}{Kronberg, P. P., Biermann, P., \&
         Schwab, F. R. 1985, \apj, 291, 693}
\reference{1}{Lehnert, M. D., Heckman, T. M., \& Weaver, K. A. 1999,
          ApJ, 523, 575}
\reference{1}{Lester, D. F., Carr, J.
         S., Joy, M., \& Gaffney 1990, \apj, 352, 544}
\reference{1}{Lynds, C. R., \& Sandage, A. R. 1963, ApJ, 137, 1005}
\reference{1}{McKeith, C. D., Greve, A., Downes, D., \& Prada, F. 1995,
               \aap, 293, 703}
\reference{1}{Nakai, N. 1989, \pasj, 41, 1107}
\reference{1}{Nakai, N., Hayashi, M., Handa, T., Sofue, Y.,
                  Hasegawa, T., \& Sasaki, M. 1987, PASJ, 39, 685}
\reference{1}{Rupke, D. S., Veilleux, S., \& Sanders, D. B. 2002, ApJ,
               570, 588}
\reference{1}{Scarrott, S. M., Eaton, N., \& Axon, D. J. 1991, \mnras,
               1991, 252, P12}
\reference{1}{Schmidt, G. D., Angel, J. R. P., \& Cromwell, R. H 1976,
               \apj, 206, 888}
\reference{1}{Shopbell, P. L., \& Bland-Hawthorn, J. 1998, ApJ, 493, 129}
\reference{1}{Sofue, Y., Reuter, H. -P., Krause, M., Wielebinski, R.,
               \& Nakai, N. 1992, ApJ, 395, 126}
\reference{1}{Strickland, D. K., Ponman, T. J., \& Stevens, I. R. 1997,
               \AA, 320, 378}
\reference{1}{Strickland, D. K., \& Stevens, I. R. 2000, \mnras,
               314, 511}
\reference{1}{Suchkov, A. A., Balsara, D. S., Heckman, T. M.,
               \& Leitherer, C. 1994, \apj, 430, 511}
\reference{1}{Tenorio-Tagle, G., \& Munoz-Tunon, C. 1998,
               \mnras, 293, 299}
\reference{1}{Tomisaka, K., \& Ikeuchi, S. 1988, ApJ, 330, 695}
\reference{1}{Veilleux, S., \& Rupke, D. S. 2002, ApJ, 565, L63}
\reference{1}{Weedman, D. W., Feldman, F. R., Balzano, V. A., Ramsey, L. W.,
               Sramek, R. A., \& Wu, C. -C 1981, ApJ, 248, 105}
\reference{1}{Yoshida, M., et al. 2000, Proc. SPIE, 4009, 10}
\reference{1}{Yun, M. S., Ho, P. T. P., \& Lo, K. Y. 1993, \apjl, 411, 17}
\reference{1}{Yun, M. S., Ho, P. T. P., \& Lo, K. Y. 1994, Nature,
               372, 530}
\end{references}
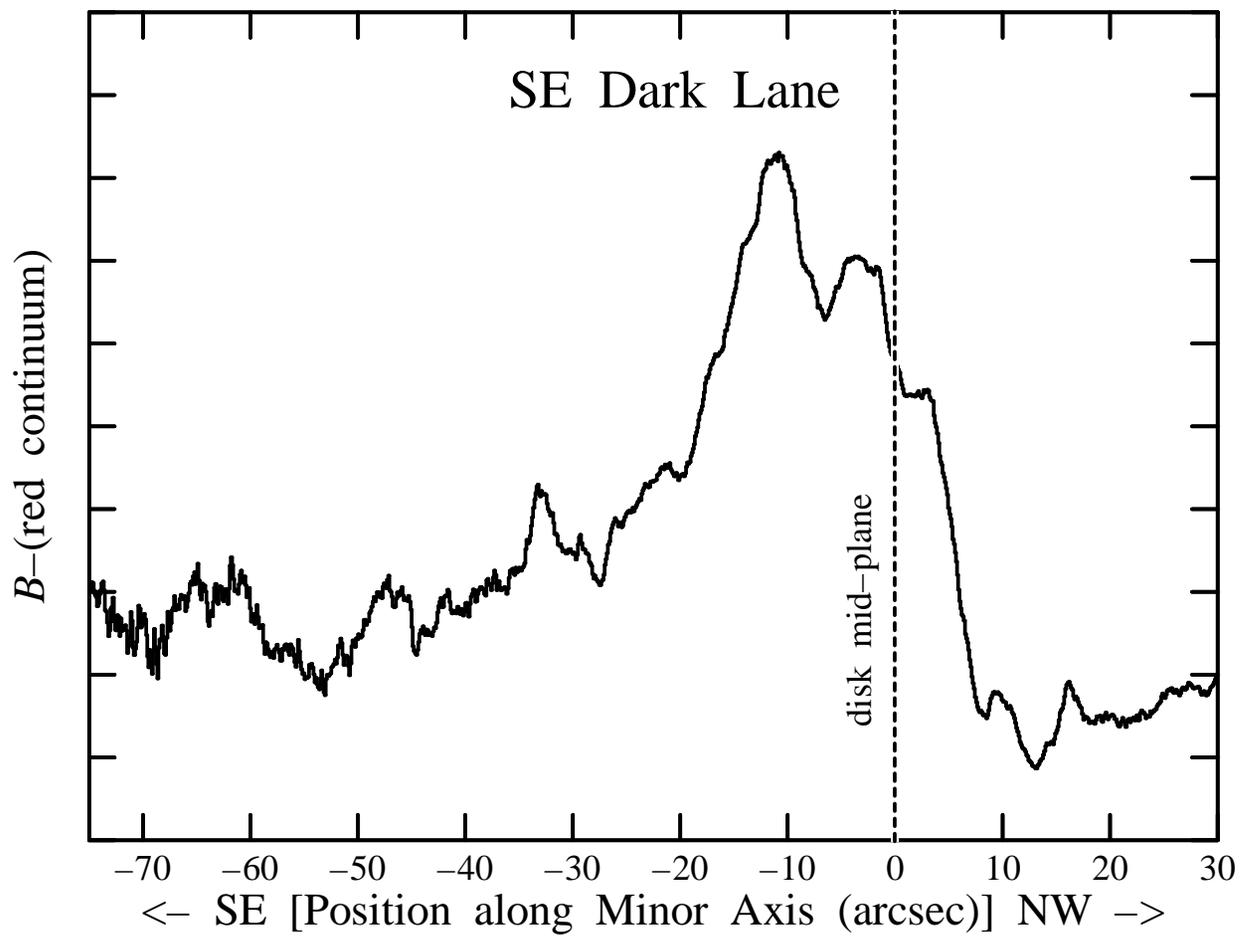
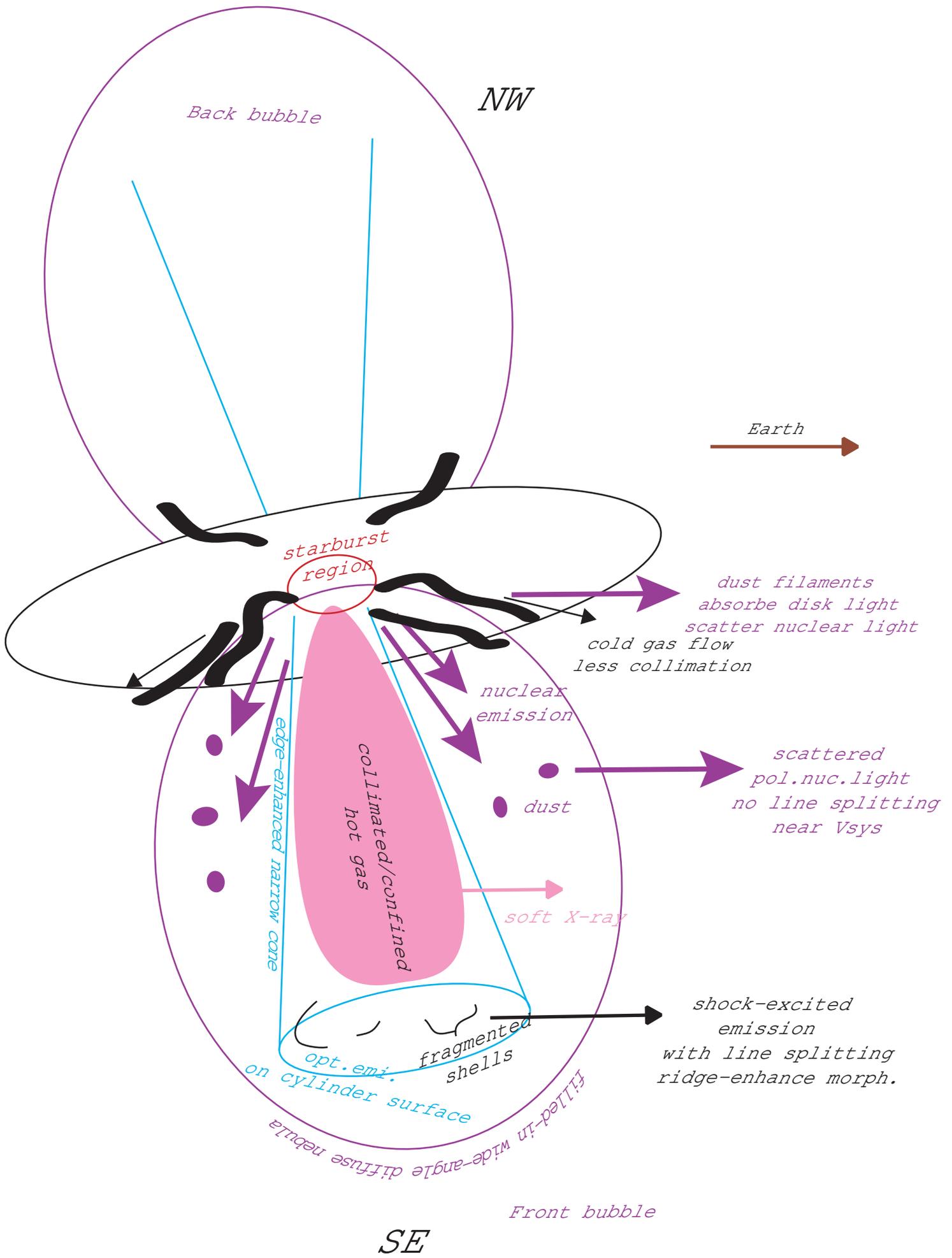
\end{document}